\documentclass[pra,twocolumn]{revtex4-1}
\usepackage{amsmath,amssymb}
\usepackage{graphicx}
\usepackage[colorlinks,linkcolor=blue,citecolor=blue,urlcolor=green]{hyperref}
\usepackage[varg]{txfonts}

\newcommand{\iu}{\mathrm{i}}
\newcommand{\dd}{\mathrm{d}}

\newcommand{\en}{\varepsilon}

\renewcommand{\vec}[1]{\mathbf{#1}}

\begin{document}

\title{Energy loss spectroscopy of Buckminster C$_{60}$ with twisted electrons:  Influence 
  of orbital angular momentum transfer on plasmon generation}

\author{Michael Sch\"uler}
\author{Jamal Berakdar}
\affiliation{Martin-Luther-University
  Halle-Wittenberg, Institute for Physics,
  Karl-Freiherr-von-Fritsch-Stra\ss{}e 3, 06120 Halle, Germany}

\begin{abstract}
  Recent experimental progress in creating and controlling singular
  electron beams that carry orbital angular momentum allows for new
  types of local spectroscopies. We theoretically investigate the
  twisted-electron energy loss spectroscopy (EELS) from the C$_{60}$
  fullerene. Of particular interest are the strong multipolar
  collective excitations and their selective response to the orbital
  angular momentum of the impinging electron beam.  Based on ab-initio
  calculations for the collective response we compute EELS signals
  with twisted electron beams and uncover the interplay between the
  plasmon polarity and the amount of angular momentum transfer.
\end{abstract}

\maketitle

\section{Introduction}

Collective excitations in nanostructures are at the heart of of the
research field nanoplasmonics~\cite{maier_plasmonics:_2005}. A
standard and widely utilized method to access the details of such
collective modes is the electron-energy loss spectroscopy (EELS)
\cite{schattschneider_fundamentals_2012}.  With the advent of vortex,
or twisted beams
~\cite{verbeeck_production_2010,van_boxem_inelastic_2015,guzzinati_probing_2016}
it is timely to explore the add-on features when performing EELS with
such beams.  A particular aspect of vortex beams is that they carry a
definite and controllable amount of orbital angular momentum (OAM)
which is related to the topological charge of the vortex.  Remarkably,
vortex beams were also realized in a transmission electron microscope
(TEM) allowing so for an atomic spatial resolution.  Typical phenomena
associated with the OAM of the twisted beam are dichroism in magnetic
systems~\cite{lloyd_quantized_2012}, and new types of Landau
states~\cite{bliokh_electron_2012,schattschneider_imaging_2014}.\\ Using
a similar concept as for the generation of the vortex
beams~\cite{grillo_holographic_2015}, the angular momentum after
scattering from the probe can be determined. Exploiting this feature
one may employ vortex-based EELS to investigate the system response
not only at a particular the linear momentum transfer, but also for a
well-defined \emph{orbital angular momentum transfer} (OAMT).  One
consequence for instance is that multipolar excitations can be
accessed even at small (linear) momentum transfer, which is known as
the optical limit in conventional EELS.

A prominent molecular example, where the excitation energy varies
significantly with the multipolarity, is the Buckminster fullerene
C$_{60}$~\cite{verkhovtsev_interplay_2012,verkhovtsev_quantum_2013,li_plasmon_2013}. In
our previous studies~\cite{schuler_disentangling_2015} we have already
developed an accurate model, based on first-principle calculations,
which is very suitable for studying EELS. In this contribution, we
employ a slightly improved version of the model with the main focus on
elucidating how the control of the OAMT can be utilized to map out
multipolar excitations. After introducing the general theoretical
formulation, we consider both, the case of an isolated molecule and a
two-dimensional film of molecules. We show that by fixing the OAMT the
encoded phase information results in specific features in the
spectra. This effect is most pronounced for spectroscopy on a single
molecule, but it also prevails for crystallized C$_{60}$. Although we
focus on the Buckminster fullerene here, the methodology and the
formula below are general and applicable to other systems.

The paper is organized as follows. In Sect.~\ref{sec:theory} we
revisit the basic formulation of EELS in view of a more general
projectile wave-functions, such as "twisted" electrons. Our
parameterization of the underlying plasmonic response of the system is
also discussed. Based on this model,
we first illustrate the control of the multipolarity in
Sect.~\ref{sec:results} by studying the vortex-based EELS from a
single molecule. After that we turn to a
crystallized surface. Atomic units are used unless stated otherwise.

\section{Theoretical formulation\label{sec:theory}}

Given the initial and the final asymptotic states of the electrons are
known and denoted by respectively $\psi_i(\vec{r})$ and
$\psi_f(\vec{r})$, the Fermi's golden rule allows for the calculation
of the transfer rate as \cite{joachain_quantum_1975}
\begin{equation}
  \label{eq:fermi1}
  \Gamma \propto \sum_{\alpha\ne 0} \left|\langle \Psi_{\alpha}, \psi_f | \hat{V}_{ee} | \Psi_0,
    \psi_i \rangle   \right|^2 \delta(E_0+ \en_i - E_\alpha - \en_f) \ ,
\end{equation}
where $|\Psi_0\rangle$ ($|\Psi_\alpha\rangle$) are the ground
(excited) states of the targets with corresponding energy $E_0$
($E_\alpha$), $\en_{i,f}$ is the energy of the incoming or outgoing
electrons, respectively, and $\hat{V}_{ee}$ is the Coulomb
interaction. In a typical EELS setup the energy of the impinging
electrons is much larger than the typical target excitations which allows
neglecting exchange effects and
simplifying the transfer rate eq.~\eqref{eq:fermi1} to
\begin{equation}
  \label{eq:fermi2}
  \Gamma \propto \sum_{\alpha\ne 0}
  \left|\langle \Psi_{\alpha}| \hat{V}_{if} | \Psi_0 \rangle
    \right|^2 \delta(E_0+ \en_i - E_\alpha - \en_f) \ .
\end{equation}
The operator $\hat{V}_{if} = \int\dd \vec{r}\, V_{if}(\vec{r})
\hat{\psi}^\dagger(\vec{r})\hat{\psi}(\vec{r})$ (expressed in
second quantization) stands for the effective potential acting on the target,
\begin{equation}
  \label{eq:vif1}
  V_{if}(\vec{r}) = \int\!\dd\vec{r}^\prime\,
  v(\vec{r}-\vec{r}^\prime) \psi^*_i(\vec{r}^\prime)\psi_f(\vec{r}^\prime) \ ,
\end{equation}
while $v(\vec{r})=1/|\vec{r}|$ is the Coulomb potential. The
fluctuation-dissipation theorem \cite{giuliani_quantum_2005} provides
a link of the expression eq.~\eqref{eq:fermi2} to the density-density response
function \cite{onida_electronic_2002} $\chi(\vec{r},\vec{r}^\prime;\omega)$ by
\begin{equation}
  \label{eq:fermi3}
  \Gamma(\omega) \propto -\int\!\dd\vec{r}\!\int\!\dd\vec{r}^\prime\, V_{if}(\vec{r})
  \mathrm{Im}[\chi(\vec{r},\vec{r}^\prime;\omega)]V^*_{if}(\vec{r}^\prime) \ .
\end{equation}
Here, $\omega = \en_i-\en_f>0$ denotes the energy loss. Alternatively
one can combine the convolution with the Coulomb potential in
eq.~\eqref{eq:vif1} with the response function by introducing the
dynamically screened interaction $W(\vec{r},\vec{r}^\prime;\omega) =
v(\vec{r}-\vec{r}^\prime) + \delta W(\vec{r},\vec{r}^\prime;\omega)$
with
\begin{equation}
  \delta W(\vec{r},\vec{r}^\prime;\omega) = \int\!\dd\vec{r}_1\!\int\!\dd\vec{r}_2 \,
  v(\vec{r}-\vec{r}_1) \chi(\vec{r}_1,\vec{r}_2;\omega) v(\vec{r}_2-\vec{r}^\prime) \ ,
\end{equation}
yielding
\begin{equation}
  \label{eq:fermi4}
  \Gamma(\omega) \propto -\int\!\dd\vec{r}\!\int\!\dd\vec{r}^\prime\, \psi^*_i(\vec{r})\psi_f(\vec{r})
  \mathrm{Im}[\delta W(\vec{r},\vec{r}^\prime;\omega)]\psi_i(\vec{r}^\prime)\psi^*_f(\vec{r}^\prime) \ .
\end{equation}
We note that eq.~\eqref{eq:fermi4} can also be derived from classical
considerations \cite{garcia_de_abajo_optical_2010}.

So far the wave function of the in- or outgoing electrons have not
been specified. Depending on the actual experimental setup, a wide
range of scenarios is possible. Here we focus on spectroscopy with
beams carrying orbital angular momentum, called twisted electron
beams. They can be described by \cite{verbeeck_new_2012}
\begin{equation}
  \varphi_{\ell k}(\vec r) = e^{\iu \ell \phi} e^{\iu k z} F_\ell(R) \ ,
\end{equation}
where cylindrical coordinates $(R,\phi,z)$ have been used. For later
convenience, we express the position vectors as $\vec r = \vec R + z
\vec{e}_z$, with $|\vec R| = R$, and $\vec{e}_z$ stands for the unit
vector in $z$ direction. Note that the radial profile $F_\ell(R)$
(which is kept general at this point) can depend on a transverse
momentum component. In the case of wide beams as compared to the
typical system size, transverse momentum transfer does, however, not
play an important role~\cite{schattschneider_comment_2013}. We will
hence omit this momentum dependence of the profiles $F_\ell(R)$. The
normalization is fixed by the orthonormality condition
\begin{equation}
  \langle \varphi_{\ell k}|\varphi_{\ell^\prime k^\prime}\rangle = \delta_{\ell \ell^\prime}
  \delta(k-k^\prime) \ .
\end{equation}
Provided such twisted electrons scatter from a target besides the
momentum in longitudinal direction, \emph{angular} momentum might be
transferred. The consequences of this effect depend on how the
outgoing electrons are detected. We now focus on two typical
scenarios.

\subsection{Conventional TEM\label{subsec:contem}}

In the TEM setup electrons are collected in a wide-angle analyzer
after being transmitted through the sample
\cite{egerton_electron_2009,garcia_de_abajo_optical_2010}. For this
reason, the angular momentum of the outgoing electrons is not
determined. Assuming the electron beam is prepared in a twisted state
$\varphi_{\ell k}(\vec r)$ and is detected with transverse momentum
$\vec{p}_\perp$, one finds for the momentum-resolved EELS signal
\begin{align}
  \frac{\dd \Gamma_\ell(\omega)}{\dd \vec{p}_\perp} &\propto
  -\int\!\dd\vec{r}\!\int\!\dd\vec{r}^\prime\,
  e^{\iu \ell(\phi^\prime-\phi)} e^{\iu \vec{p}_\perp \cdot(\vec R-\vec{R}^\prime)}
  e^{\iu\omega(z^\prime-z)/k} \nonumber \\ &\quad \times
  F^*_{\ell}(R)F_{\ell}(R^\prime)\mathrm{Im}[\delta W(\vec{r},\vec{r}^\prime;\omega)]
\end{align}
Here, we have approximated the longitudinal momentum transfer $q$ by
$q=\omega/k$, which is obtained by a first-order Taylor expansion in
$\omega/\en_i$. The angular momentum of the twisted beam thus directly
influences the EELS signal provided the angular distribution is
recorded. Integrating over all possible detection directions on the
other hand,
\begin{align*}
  \Gamma_\ell(\omega) = \int\!\dd \vec{p}_\perp\,  \frac{\dd \Gamma(\omega)}{\dd \vec{p}_\perp} \ ,
\end{align*}
yields the total cross section \cite{garcia_de_abajo_optical_2010}
\begin{align}
  \label{eq:gammanotwist}
  \Gamma_\ell(\omega) &\propto - \int\!\dd \vec{R}
  \!\int^\infty_{-\infty}\!\dd z\!\int^\infty_{-\infty}\!\dd z^\prime e^{\iu\omega(z^\prime-z)/k}
  |F_\ell(R)|^2 \nonumber \\ &\quad \times
  \mathrm{Im}[\delta W(\vec R+z \vec{e}_z,\vec R + z^\prime \vec{e}_z;\omega)] \ .
\end{align}
An important conclusion to be drawn from eq.~\eqref{eq:gammanotwist}
is that the influence of the angular momentum  on
the EELS spectra enters through the  radial beam profile (that depends on $\ell$).
This is the case
 if the signal is collected by integrating over all
possible directions.

\subsection{Detection of angular momentum\label{subsec:detangmom}}

The situation changes if the angular momentum of the scattered
electrons is detected explicitly. Experimentally, this can achieved by
a holographic vortex filter that the scattered beam traverses, thus
separating the different angular momentum components by the
propagation direction~\cite{schachinger_towards_2016}.
In this case the OAMT $\Delta \ell = \ell-\ell^\prime$ becomes an
important control parameter~\cite{lloyd_quantized_2012}. Note that
this characterization is only possible if the respective axis of in-
and outgoing beam both coincide. In general, this is an approximation
which is adequate for targets smaller than the beam
waist~\cite{schattschneider_comment_2013}.
Based in thus assumption can most conveniently compute the effective
potentials by solving Poisson's equation,
\begin{equation}
  \label{eq:poisson3d}
  \nabla^2 V_{if}(\vec r) = -4\pi \varphi^*_{\ell k}(\vec r)
  \varphi_{\ell^\prime k^\prime}(\vec r)  \ ,
\end{equation}
exploiting the cylindrical symmetry. The ansatz $V_{if}(\vec r) =
e^{\iu \Delta \ell \phi} e^{\iu q z} w_{\ell,\ell^\prime}(q;R)$ (where
$q=k-k^\prime$) reduces eq.~\eqref{eq:poisson3d} to the radial
Poisson equation
\begin{equation}
  \label{eq:radpoisson}
  \left[\frac{\dd^2}{\dd R^2}+ \frac{1}{R}\frac{\dd}{\dd R} +
    \frac{\Delta \ell^2}{R^2} - q^2\right]w_{\ell,\ell^\prime}(q;R) =
  -4\pi F_{\ell}(R) F_{\ell^\prime}(R) \ ,
\end{equation}
which is solved in terms of the its Green's function
$g_{mq}(R,R^\prime) = \pi(2-\delta_{m,0})I_m(q R_<) K_m(q R_>)$. Here,
$I_m(x)$ and $K_m(x)$ denote the modified Bessel functions of first
and second kind, respectively. As usual,
$R_<=\mathrm{min}(R,R^\prime)$ and $R_>=\mathrm{max}(R,R^\prime)$. 
For the radial part of the potential one finds
\begin{equation}
  \label{eq:radpoissiongreen}
  w_{\ell,\ell^\prime}(q;R) =
  -4 \pi \int^\infty_0\!\dd R^\prime\, R^\prime \, g_{|\ell-\ell^\prime| q}(R,R^\prime)
  F^*_{\ell}(R^\prime) F_{\ell^\prime}(R^\prime) \ .
\end{equation}
For a target possessing almost perfect spherical symmetry such as the
C$_{60}$ molecule, the expansion of the density-density response
function in terms of fluctuation densities reads
\begin{equation}
  \label{eq:chisphere}
  \chi(\vec r,\vec{r}^\prime;\omega) = \sum_{\nu L M}\xi_{\nu
    L}(\omega) \rho_{\nu L}(r) \rho_{\nu L}(r^\prime)
  Y^*_{LM}(\hat{\vec r}) Y_{LM}(\hat{\vec r}^\prime) \ .
\end{equation}
Therefore, eq.~\eqref{eq:fermi3} attains the form
\begin{align}
  \label{eq:gammaspherical}
  \Gamma_{\ell \ell^\prime}(\omega) \propto -\sum_{\nu L M} \mathrm{Im}[\xi_{\nu L}(\omega)]
  \left|\int\!\dd \vec{r}\, \rho_{\nu L}(r) Y^*_{LM}(\hat{\vec r}) V_{if}(\vec r)\right|^2
\end{align}
An important special case occurs if the beam axis points through
the center of the C$_{60}$ molecule, as the integration over the angle
$\phi$ is simplified by
\begin{align}
  \label{eq:selrule}
  \int^{2\pi}_0\!\dd \phi\, Y^*_{LM}(\hat{\vec r}) e^{\iu \Delta \ell
    \phi} = 2\pi\widetilde{P}^M_L(\cos\theta) \delta_{\Delta \ell,M} \ .
\end{align}
Here, $\widetilde{P}^M_L(x)$ stands for the associated Legendre
polynomials normalized in accordance with the spherical harmonics. The
selection rule $\Delta \ell = M$ limits the sum over $L$ by $L\ge
|\Delta \ell|$ in eq.~\eqref{eq:gammaspherical} and hence excludes certain
multipolar modes.

\begin{figure}[t]
  \includegraphics[width=0.9\columnwidth]{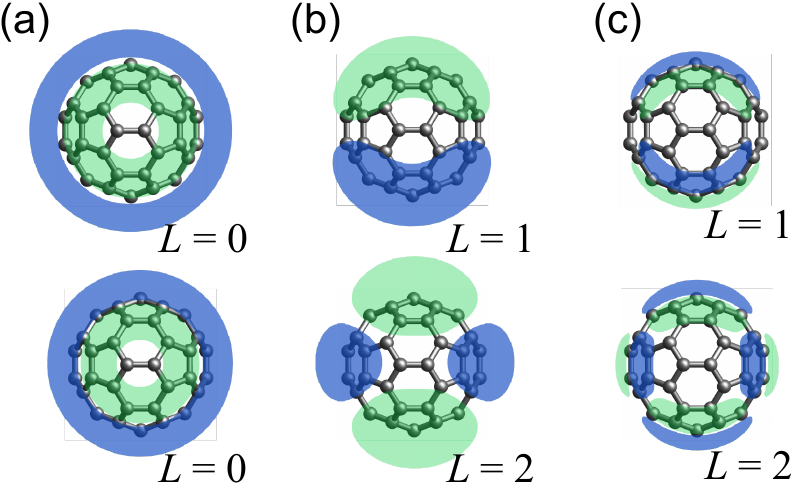}
  \caption{(Color online) Illustration of the plasmon modes by the
    area in the $x$-$z$ plane where the model fluctuation densities
    $\rho_{\nu L,M=0}(\vec{r})>0$ (light green) and $\rho_{\nu
      L,M=0}(\vec{r})<0$ (darker blue). (a) Volume plasmons V1 (upper)
    and V2 (lower plot), (b) symmetric surface (SS) plasmons for
    $L=1,2$, and (c) anti-symmetric surface (AS) modes with $L=1,2$.
    \label{fig:plmodes} }
\end{figure}

\subsection{Density-density response of the C$_{60}$ molecule
\label{sec:ddresponse}}

The central ingredient determining the (vortex) EELS signal is the
density-density response function of the system, which comprises all
types of excitations present in the system. This includes plasmons and
particle-hole (\mbox{$p$--$h$}) excitations. Qualitative insights on
the collective density fluctuations can be gained from semi-classical
considerations~\cite{verkhovtsev_formalism_2012,bolognesi_collective_2012},
where the electronic density C$_{60}$ molecule is approximated by a
spherical shell. The density can thus only fluctuate at the inner and
the outer surface, respectively, giving rise to symmetric or
anti-symmetric oscillations and a volume plasmon. Our parameterization
from ref.~\cite{schuler_disentangling_2015} provides an accurate fit
to the fully-fledged first-principle calculations based o
time-dependent density-functional theory (TDDFT) and yet allows for an
intuitive classification of the plasmon modes as given in the
semi-classical model. In particular, the density-density response
function $\chi(\vec{r},\vec{r}^\prime;\omega)$ is expanded as in
eq.~\eqref{eq:chisphere}, identifying the index $\nu$ with the radial
quantum numbers. We distinguish between symmetric surface (SS)
plasmons characterized by $\nu=\mathrm{SS}$ and multipolarity $L\ge
1$, anti-symmetric surface (AS) plasmons ($\nu=\mathrm{AS}, L\ge 1$)
and two types of volume plasmons ($\nu=\mathrm{V1},\mathrm{V2},
L=0$). The quenching of the volume plasmons (which is lacking in the
semi-classical shell model) is a consequence of the delocalized nature
of electron density. The plasmon modes entering our model are sketched
(up to $L=2$) in Fig.~\ref{fig:plmodes}.

\begin{figure}[t]
  \includegraphics[width=\columnwidth]{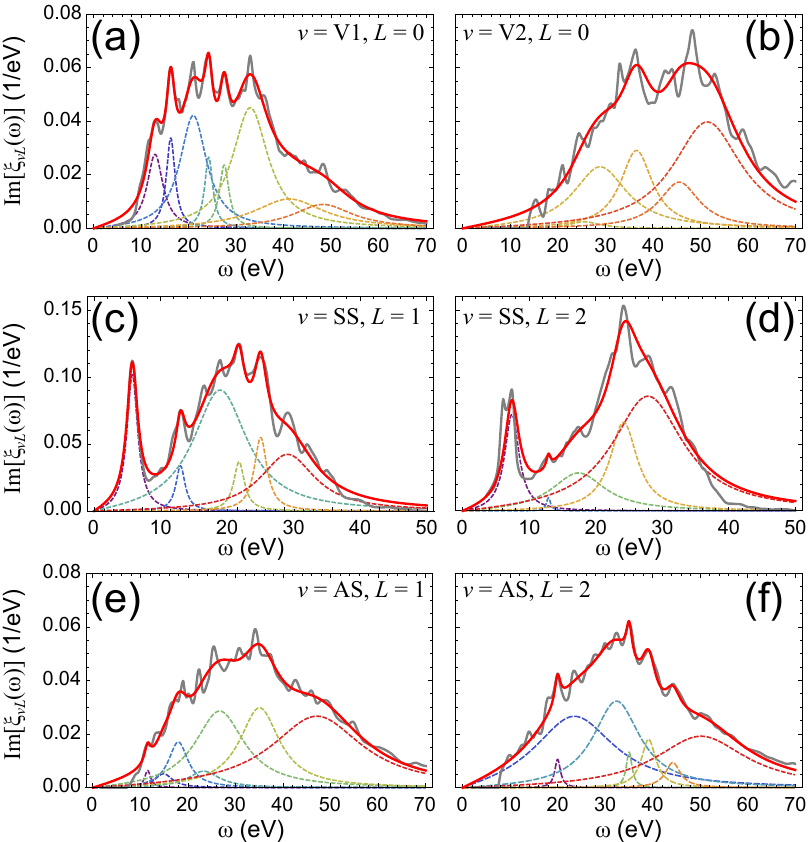}
  \caption{Multi-peak fits of the frequency dependence
    $\xi_{\nu L}(\omega)$ of the density-density response function
    (cf. eq.~\eqref{eq:chisphere}) in terms of the volume ((a) and (b)),
    the symmetric surface ((c) and (d)), and the anti-symmetric surface ((e)
    and (f)) plasmons. Solid gray lines: TDDFT results
    $\xi^\mathrm{TDDFT}_{\nu L}(\omega)$ from
    ref.~\cite{schuler_disentangling_2015}; solid red lines: fitted
    frequency dependence $\xi^\mathrm{fit}_{\nu L}(\omega)$; dashed
    lines: individual peak contributions (see
    eq.~\eqref{eq:multipeak}).
    \label{fig:multipeak} }
\end{figure}

The model from ref.~\cite{schuler_disentangling_2015} is constructed from
fitting functions for the spectra $\xi_{\nu L}(\omega)$ and the
fluctuation densities $\rho_{\nu L}(r)$, which allowed for an accurate
modeling of the full density-density response
function~\cite{schuler_electron_2016}. Here, we use an improved
version of the fitting procedure for the frequency dependence. Taking
the spectral functions $\xi^\mathrm{TDDFT}_{\nu L}(\omega)$ from our
TDDFT calculations, we select the dominant peaks assuming a generic
form
\begin{align}
  \label{eq:multipeak}
  \xi^\mathrm{TDDFT}_{\nu L}(\omega) \approx \xi^\mathrm{fit}_{\nu L}(\omega)
  = \sum^{N_{\nu L}}_{i=1} A_{i \nu L}
\frac{2\Omega_{i \nu L}}{(\omega+\iu \Gamma_{i \nu L})^2 -
  \Omega^2_{i \nu L}} \ .
\end{align}
The weights $A_{i \nu L}$, peak frequencies $\Omega_{i \nu L}$, and the
broadening $\Gamma_{i \nu L}$ are then obtained from a least-square
fit. The obtained spectra (up to $\ell=2$) are compared to the results
of the \emph{ab initio} calculations from
ref.~\cite{schuler_disentangling_2015} in Fig.~\ref{fig:multipeak}.

\section{Results\label{sec:results}}

With the general theoretical formulation from Sect.~\ref{sec:theory}
and an accurate model for the fluctuation densities and the
corresponding spectra at hand, we can now analyze the inelastic
scattering of the vortex beams from the C$_{60}$ molecule.
In line of a typical experimental
realization~\cite{bliokh_electron_2012,van_boxem_inelastic_2015,schachinger_peculiar_2015},
we choose Laguerre-Gauss modes as an approximation to the profile $F_\ell(R)$ of
the vortex beams:
\begin{equation}
  \label{eq:profile}
  F_\ell(R)= \frac{2}{\sqrt{\ell !}W_0} \left(\frac{\sqrt{2}R}{W_0}\right)^{|\ell|}
  e^{-(R/W_0)^2} L^p_{|\ell|}(2 (R/W_0)^2) \ .
\end{equation}
Here, $L^p_{|\ell|}(x)$ denote the associated Laguerre polynomials. We
fix the radial node number to be $p=0$. Note that $\varphi_\ell(\vec
r)$ is not an eigenstate of the free-particle Hamiltonian in this
case. However, as the energy is carried in the longitudinal ($z$)
direction, the energy of the beam is still sharply
defined. Representing the vortex beams by the Laguerre-Gauss
profile~\eqref{eq:profile} shifts the dependence on the transverse
momentum to the beam waist $W_0$. We assume that $W_0$ is preserved
upon scattering -- an approximation that relies on the small size of
the molecules on the scale of $W_0$. By varying $W_0$ of the outgoing
beam we confirmed that this assumption is justified to very good
accuracy. This can be understood by the weak dependence of the
effective potential on the beam profile, which is discussed now.

\subsection{Effective potential}

\begin{figure}[t]
  \includegraphics[width=\columnwidth]{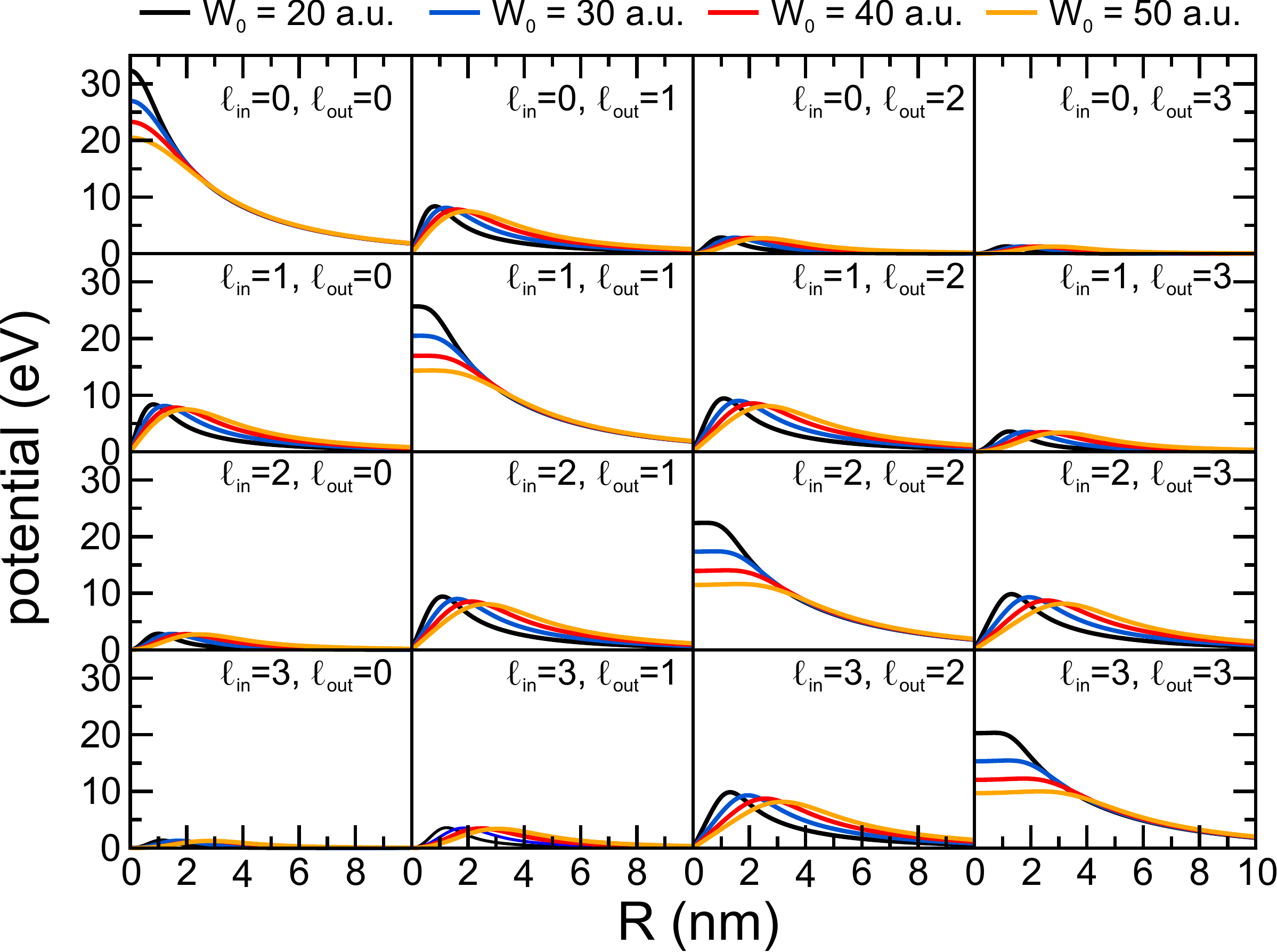}
  \caption{The radial part $w_{\ell_\mathrm{in}, \ell_\mathrm{out}}(q;R)$ of the
    effective potential for the typical value $q=0.01$~a.\,u.
    and $0\le \ell_\mathrm{in} \le 3$, $0\le \ell_\mathrm{out} \le 3$. \label{fig:Veff} }
\end{figure}

In fig.~\ref{fig:Veff} we present the radial part of the effective
potential $V_{if}(\vec r)$ as discussed in
subsection~\ref{subsec:detangmom} for typical values of the beam waist
$W_0$. As we can infer from fig.~\ref{fig:Veff}, the effective potential
quickly drops with increasing OAMT $\Delta \ell =
\ell_\mathrm{in}-\ell_\mathrm{out}$, which is explained by the
decreasing overlap of the respective beam profiles. The potential
displays a plateau behavior around $R=0$ for
$\ell_\mathrm{in}=\ell_\mathrm{out}$, while it vanishes at this point
for $\ell_\mathrm{in}\ne\ell_\mathrm{out}$. The asymptotic behavior is
determined by $e^{-q R}/\sqrt{q R}$, i.\,e. for small momentum
transfer as typically encountered in high-energy EELS, the effective
potential can be quite long-ranged affecting the molecules situated
far away from the beam axis. This is very different from photons
carrying orbital angular momentum. In a conventional EELS setup, the
effective potential reads $4\pi/q^2 e^{\iu \vec{q}\cdot \vec r}$ and
thus exhibits a quadratic divergence for $q\rightarrow 0$. The
effective potential caused by scattering of twisted electrons on the
other hand shows a logarithmic divergence, as expected for a
two-dimensional regularization due to using beams with a finite width.

\subsection{Loss spectra: beam focused on molecule\label{subsec:singlemol}}

If the beam axis  passes through the molecule's center the
selectivity with respect to the OAMT is most pronounced due to
eq.~\eqref{eq:selrule}. As the energy of impinging electrons is large
($\en_i = 60$~keV), the longitudinal momentum transfer is in the range
of $q\simeq \omega / \sqrt{2\en_i} \lesssim 0.03$~a.\,u. approaching
the optical limit for the C$_{60}$ molecule. As previously discussed
\cite{schuler_disentangling_2015}, the dominant excitations for small
$q$ are the SS plasmons and the dipole SS plasmon
in particular. Volume plasmons can however also be induced due to the
radial dependence of the beam. Increasing the OAMT $|\Delta \ell|$ we
expect the loss spectra are shifted to higher energies, as the
frequencies of the SS plasmons grow with the multipolarity.

To  confirm this dependence, we computed the loss spectra for
different values of $\ell_\mathrm{in}$ and fixed the angular momentum
of the scattered electrons at $\ell_{out}=0$. The OAMT to the system is
thus $\Delta \ell = \ell_\mathrm{in}$.

For evaluating eq.~\eqref{eq:gammaspherical} we use the radial
fluctuation densities $\rho_{\nu L}(r)$ and plasmon spectra $\xi_{\nu
  L}(\omega)$ from ref.~\cite{schuler_disentangling_2015}. After
solving the radial Poisson equation \eqref{eq:radpoisson} by
eq.~\eqref{eq:radpoissiongreen}, the effective potential $V_{if}(\vec
r)$ is projected on the spherical harmonics with respect to the
molecule's center. Finally, the remaining integration over the distant
from the center $r$ is performed.  The momentum transfer is replaced
by $q=\omega/k$.

\begin{figure}[t]
  \includegraphics[width=0.9\columnwidth]{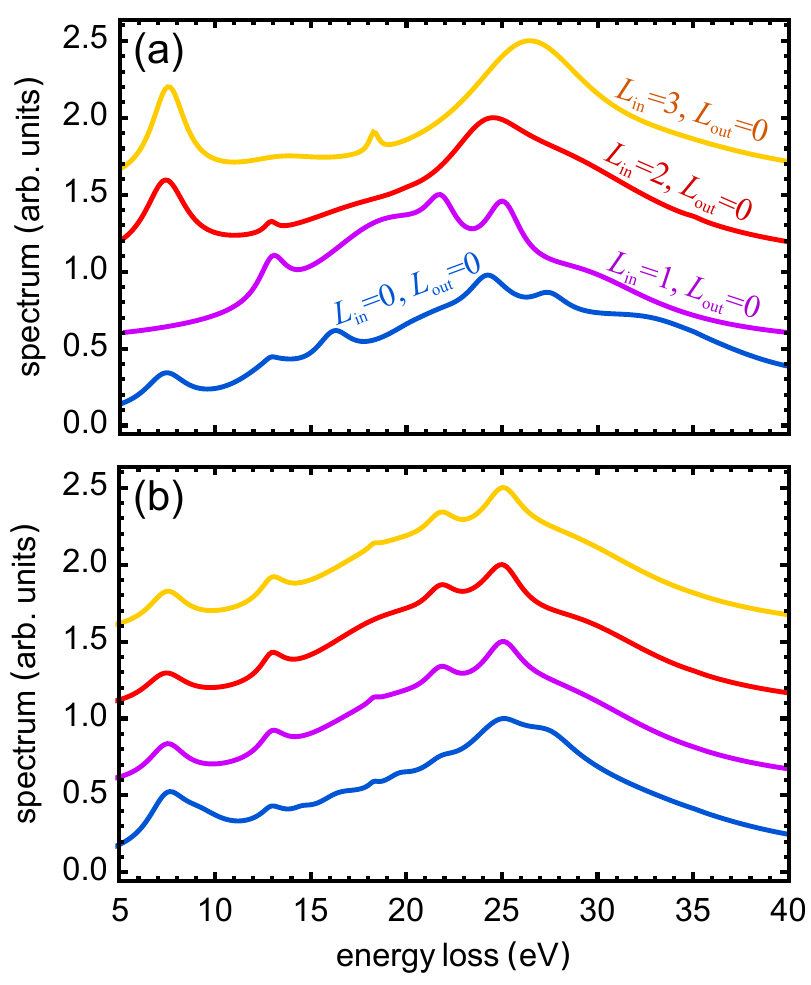}
  \caption{Twisted electron energy loss spectra when beam axis coincides
   with the center
    of the C$_{60}$ molecule (beam waist $W_0=30$~a.\,u.). (a) Fixed
    outgoing angular momentum $L_\mathrm{out}=0$ for different angular
    momenta $L_\mathrm{in}$ of the impinging beam. (b) Loss spectra
    for the same incoming angular momenta as in (a), but without fixing
    $L_\mathrm{out}$. The spectra have been normalized and shifted for
    a better visibility. \label{fig:spectra1} }
\end{figure}

The resulting normalized
spectra are presented in Fig.~\ref{fig:spectra1}(a). For
$\ell_\mathrm{in}=\ell_\mathrm{out}=0$, only volume plasmons can be
excited, leading to a broad loss spectrum which is consistent with the
frequency dependence in Fig.~\ref{fig:multipeak}(a). For
$\Delta \ell=1$, dipole plasmons can be induced (predominantly the SS
plasmon). The loss spectrum is therefore similar to the optical
absorption spectrum~\cite{reinkoster_photoionization_2004}. Increasing
the OAMT, the plasmon dispersion with respect to the multipolarity leads
to a shift of the spectra to higher energies. Furthermore, due to the
changed beam profile, AS plasmons can also be induced, which further
shifts the spectra. The dependence of the loss spectra on the angular
momenta is quantified in table~\ref{tab:w0mx}, where we give the
overall peak positions (obtained by a Lorentzian fit) as a function of
$\ell_\mathrm{in}$ and $\ell_\mathrm{out}$.

The situation changes drastically if the requirement of detecting the
outgoing angular momentum is dropped. As elaborated upon in
subsection~\ref{subsec:contem}, the effect of the OAMT should
diminish. This is indeed consistent with our results for this case
(Fig.~\ref{fig:spectra1}(b)). The loss spectra exhibit a very weak
dependence on the initial angular momentum of the beam which arises
due to a changed beam profile only. Hence, not detecting the
angular momentum leads to a loss of  phase information which is
directly reflected in the featureless spectra.

\begin{table}[b]
  \caption{Energy losses $\omega$ where the cross section is peaked, computed by
    performing a fit by a single Lorentzian in the peak region.
    The beam has a waist of $W_0=30$~a.\,u. and is aligned with the molecule's center.
    Values are shown in eV. \label{tab:w0mx}}
\begin{ruledtabular}
  \begin{tabular}{cc|c|c|c|c|c}
      & $\ell_\mathrm{in}$ & 0 & 1 & 2 & 3 & 4 \\
   $\ell_\mathrm{out}$ & & & & & \\
   \hline
   0& & 	26.29 &	21.88&	25.55&	26.81& 27.07\\
   1& & 	21.88 &	24.97&	21.87&	25.53& 26.74\\
   2& & 	25.55 &	21.87&	24.89&	21.87& 25.53\\
   3& & 	26.81 &	25.53&	21.87&	24.85& 21.87\\
   4& & 	27.07 &	26.74&	25.53&	21.87& 24.82\\
  \end{tabular}
\end{ruledtabular}
\end{table}

\subsection{Loss spectra: crystalline phase}

\begin{figure}[t]
  \includegraphics[width=\columnwidth]{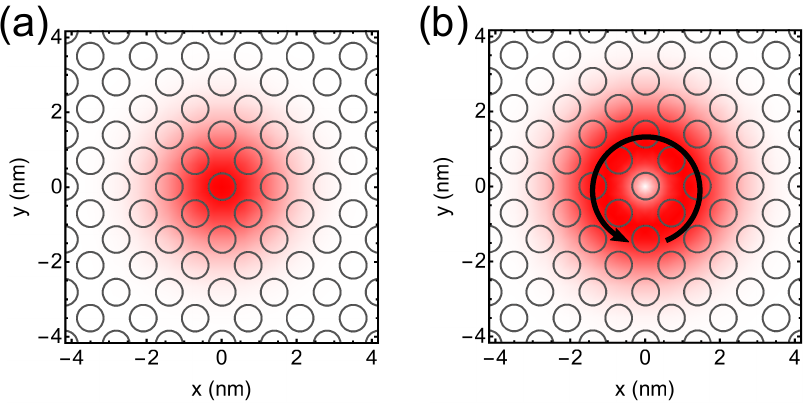}
  \caption{Illustration of the (111) surface layer of C$_{60}$
    molecules (gray circles) along with the beam profile $F_{\ell}(R)$
    (density plot) for (a) $\ell=0$ and (b) $\ell=1$. The direction of
    the phase variation is indicated by the black
    arrow. \label{fig:geom}}
\end{figure}

Conducting an EELS experiment on isolated C$_{60}$ is very
challenging, as preparing single molecules on the substrate used in
the TEM setup is hardly possible. It is much more likely that the
fullerenes crystallize on the surface of the substrate, forming a few
layers of an FCC crystal (lattice constant
$a(\mathrm{C}_{60}) = 1.4154$~nm at room temperature). To
describe this setup theoretically, based on the previously employed
model, we assume that the individual contributions of the molecules
can be summed to obtain an adequate approximation to the response of
the crystal:
\begin{align}
  \label{eq:gammasurf}
  \Gamma^\mathrm{crys}_{\ell \ell^\prime}(\omega) \propto - \sum_{n\in\mathrm{latt}}
  \sum_{\nu L M} &\mathrm{Im}[\xi_{\nu L}(\omega)] \\
  &\times\left|\int\!\dd \vec{r}\, \rho_{\nu L}(r) Y^*_{LM}(\hat{\vec r})
  V_{if}(\vec{r}+\vec{r}_n)\right|^2 \ . \nonumber
\end{align}
Here, $\vec{r}_n$ denotes the lattice sites.  This treatment ignores the
hybridization of the plasmon modes into corresponding bands. The
significance of such effects is not completely understood at the
moment; first calculations \cite{koval_computation_2015} report quite
similar spectra as compared to the gas
phase~\cite{bolognesi_collective_2012}.
As thin films are best suited
for TEM experiments, we consider a single layer of molecules here. The
geometry of the C$_{60}$ is compared to the typical beam extensions in
Fig.~\ref{fig:geom}

Note that the beam axis does not pass through most of the molecules'
centers, resulting in less sharply defined OAMT (which is defined with
respect to the beam). The scenario of decentered beams questions the
assumption of keeping the beam axis and waist $W_0$ constant
throughout the scattering process. For the molecules located not to
far from the vortex center, including the area of maximum intensity
(which has the largest contribution to the total signal), the validity
of this approximation has been underpinned in
ref.~\cite{schattschneider_is_2014}.

\begin{figure}[t]
  \includegraphics[width=0.9\columnwidth]{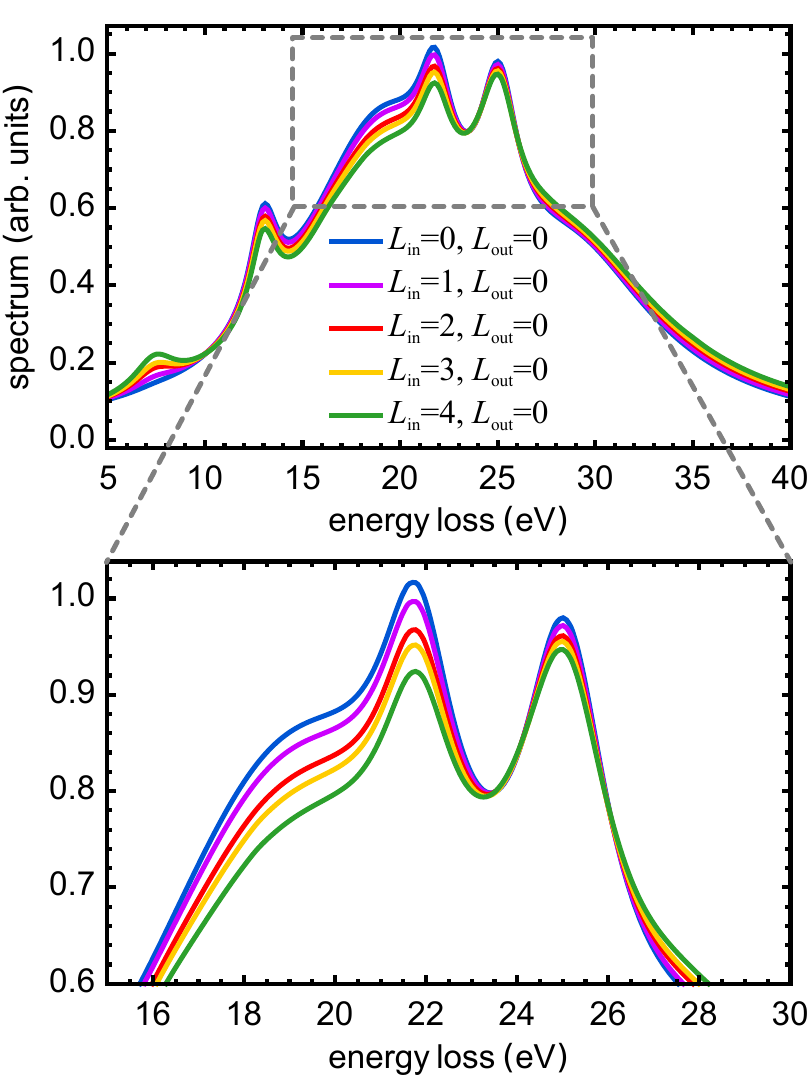}
  \caption{Loss spectra for scattering from one layer of FCC C$_{60}$
    ($W_0=30$~a.\,u.) for different values of the OAM of the ingoing
    beam $L_\mathrm{in}$, and fixed outgoing OAM $L_\mathrm{out}=0$
    (geometry as in Fig.~\ref{fig:geom}). The curves have been
    normalized to the same frequency-integrated value. The region in
    the dashed rectangle is magnified in the lower
    panel.  \label{fig:spectra2}}
\end{figure}

Analogously to the single-molecule case, we first analyze the scenario
where the angular momentum of the outgoing electrons is explicitly
detected. Evaluating the cross section~\eqref{eq:gammasurf} yields the
loss spectra presented in Fig.~\ref{fig:spectra2}. First we note that
the difference between the spectra, when varying the OAMT, is not as
pronounced as for the single molecules. This is a result of the
collective response of many molecules located off-center with respect
to the beam axis, as the OAM (which depends on the reference
coordinate system) of the beam is blurred when considered from the fullerenes'
point of view. Hence, in spite the total OAMT is fixed, off-center
molecules experience different many OAM components. Hence, the
multipolar excitations can not be controlled as efficiently as
before. Nevertheless, the phase information encoded in the OAM of the
vortex beam leads to notable differences in the loss
spectra. Generally, the trend is as in
subsection~\ref{subsec:singlemol}: the increased probability to induce
multipolar excitations with growing OAMT shifts the spectra to higher
frequencies. This behavior is mostly reflected in the less and less
pronounced shoulder at $\omega\simeq 19$~eV (see the zoom in the lower
panel in Fig.~\ref{fig:spectra2}), which corresponds to the dominant
feature of the SS dipole plasmon (see
Fig.~\ref{fig:multipeak}(c)). Hence, the contribution of the dipolar
plasmons is suppressed. Interestingly, the peak $\omega\simeq 7.5$~eV
is enhanced with increasing OAMT, as well. This peak corresponds to
series of particle-hole excitations \mbox{$p$--$h$} in the bound-state
manifold~\cite{usenko_femtosecond_2016}. The increasing weight of
\mbox{$p$--$h$} excitations as compared to the plasmons is a signature
of a more inhomogeneous driving acting on the system, as collective
excitations only exist at small wave vectors (small angular momenta,
respectively). This effect can also be observed in
Fig.~\ref{fig:spectra1}. In future works, we will map out the
\mbox{$p$--$h$} excitations induced by vortex beams based on our
\emph{ab initio} approach from ref.~\cite{usenko_femtosecond_2016}.

\begin{figure}[h]
  \includegraphics[width=0.8\columnwidth]{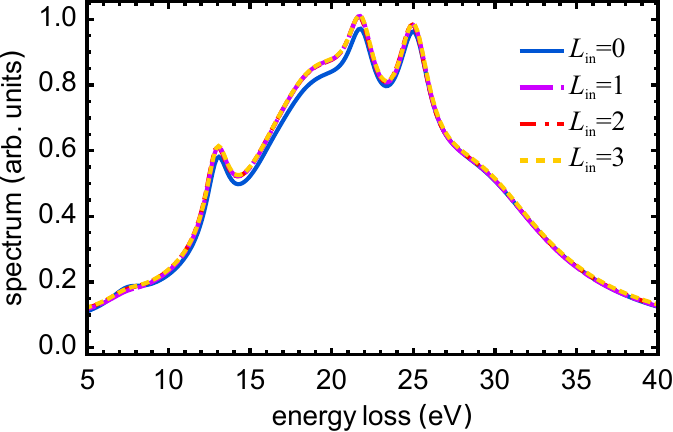}
  \caption{Loss spectra for scattering from one layer of FCC C$_{60}$
    as in Fig.~\ref{fig:spectra2}, but without fixing the outgoing OAM
    $L_\mathrm{out}$.\label{fig:spectra3}}
\end{figure}

To demonstrate that the modification of the spectra in
Fig.~\ref{fig:spectra2} depends on the phase of the vortex beam, we
recomputed the loss spectra assuming that the OAM of the outgoing
electrons is not detected (Fig.~\ref{fig:spectra3}). Analogously to
the discussion in subsection~\ref{subsec:singlemol}, we find that the
spectra for different ingoing OAM $L_\mathrm{in}$ are basically
identical, except for the case $L_\mathrm{in}=0$. The latter is due to
a quite different beam profile (see Fig.~\ref{fig:geom}(a)). Hence, it
is truly the OAM of the vortex beams (which is pure phase effect) that
can induce multipolar excitations and thus give rise to specific
features in the loss spectra.

\section{Conclusions}

We presented the theoretical description of twisted electron energy
loss spectra for both cases, when (i) the scattered electrons are
detected in the full solid angle, and (ii) when the angular momentum
of the scattered electrons is detected. While for (i) the angular
momentum of the beam, encoded in the phase has no influence, we showed
that it plays an important role in case (ii), particularly if the beam
is aligned with the single molecule center.  We applied the developed
theory for EELS with twisted electrons for fixed molecules and for a
single layer of crystallized fullerenes. The numerical findings are in
line with the formal expectations: measuring the outgoing OAM of the
beam allows controlling the OAMT and thus the multipolar
excitations. This is directly reflected in the loss spectra. In
contrast, detecting only the energy of the scattered electrons leads
to almost identical spectra with varying OAM, since the phase
information is lost and only the varying beam profile influences the
overall spectra. So, we advocate the OAM-resolved vortex-based EELS as
a powerful technique to access new information on the system's
excitations, particularly those of multipolar character.

\section*{Acknowledgments}
We are indebted to Thomas Schachinger and Michael St\"oger-Pollach for
fruitful discussions and important insights from the experimental
point of view, as well as to Yaroslav Pavlyukh for numerous
discussions.  The financial support by the grants of the German
Research Foundation (DFG) SFB762 and the Priority Programme 1840
"Quantum Dynamics in Tailored Laser Fields" (QUTIF) is gratefully
acknowledged.

%
\end{document}